\documentclass[a4paper,conference]{IEEEtran}

\usepackage[pdftex]{graphicx}

\usepackage{cite}

\usepackage{epstopdf}

\usepackage{lipsum}

\usepackage{amsmath}
\usepackage{amssymb}

\usepackage{url}

\ifCLASSOPTIONcompsoc
  \usepackage[caption=false,font=normalsize,labelfont=sf,textfont=sf]{subfig}
\else
  \usepackage[caption=false,font=footnotesize]{subfig}
\fi

\usepackage[dvipsnames]{xcolor}
\usepackage{multirow}

\usepackage{footmisc}

\usepackage[normalem]{ulem}

\newcommand{\FH}[1]{#1} 

\newcommand{\name}{SCDS}

\newcommand{\thilofeat}{CNN-T}
\newcommand{\zissfeat}{CNN-V}

\makeindex

\begin{document}

\title{Speaker Clustering Using Dominant Sets}

\author{\IEEEauthorblockN{\hspace{0.2cm}Feliks Hibraj\footnotemark{*}}
\IEEEauthorblockA{
Ca' Foscari University\\
Venice, Italy\\
feliks.hibraj@gmail.com
}\and
\IEEEauthorblockN{\hspace{0.2cm}Sebastiano Vascon\footnotemark*}
\IEEEauthorblockA{
Ca' Foscari University\\
Venice, Italy\\
sebastiano.vascon@unive.it
}
\and
\IEEEauthorblockN{Thilo Stadelmann}
\IEEEauthorblockA{ZHAW Datalab\\
Winterthur, Switzerland\\
stdm@zhaw.ch}
\and
\IEEEauthorblockN{Marcello Pelillo}
\IEEEauthorblockA{Ca' Foscari University\\
Venice, Italy\\
pelillo@unive.it}
}

\maketitle

\begin{abstract}\\

Speaker clustering is the task of forming speaker-specific groups based on a set of utterances. In this paper, we address this task by using Dominant Sets (DS). DS is a graph-based clustering algorithm with interesting properties that fits well to our problem and has never been applied before to speaker clustering. We report on a comprehensive set of experiments on the TIMIT dataset against standard clustering techniques and specific speaker clustering methods. Moreover, we compare performances under different features by using ones learned via deep neural network directly on TIMIT and \FH{other ones extracted from} a pre-trained VGGVox net. To asses the stability, we perform a sensitivity analysis on the free parameters of our method, showing that performance is stable under parameter changes. The extensive experimentation carried out confirms the validity of the proposed method, reporting state-of-the-art results under three different standard metrics. We also report reference baseline results for speaker clustering on the entire TIMIT dataset for the first time.

\end{abstract}

\makeatletter{\renewcommand*{\@makefnmark}{}
\footnotetext{* = Equal contribution}\makeatother}

\IEEEpeerreviewmaketitle

\section{Introduction}

\emph{Speaker clustering} (SC) is the task of identifying the unique speakers in a set of audio recordings (each belonging to exactly one speaker) without knowing who and how many speakers are present altogether \cite{beigi2011fundamentals}. Other tasks related to speaker recognition and SC are the following:
\begin{itemize}
\item \emph{Speaker verification} (SV): A binary decision task in which the goal is to decide if a recording belongs to a certain person or not.
\item \emph{Speaker identification} (SI): A multiclass classification task in which to decide to whom out of $n$ speakers a certain recording belongs.
\end{itemize}
SC is also referred to as \emph{speaker diarization} when a single (usually long) recording involves multiple speakers and thus needs to be automatically segmented prior to clustering. Since SC is a completely unsupervised problem (the number of speakers and segments per speaker is unknown), it is straightforward to note that it is considered of higher complexity with respect to both SV and SI. The complexity of SC is comparable to the problem of image segmentation in computer vision, in which the number of regions to be found is typically unknown.

The SC problem is of importance in the domain of audio analysis due to many possible applications, for example in lecture/meeting recording summarization \cite{anguera2012speaker}, as a pre-processing step in automatic speech recognition, or as part of an information retrieval system for audio archives \cite{ajmera2003robust}. Furthermore, SC represents a building block for speaker diarization \cite{shum2012use}.

The SC problem has been widely studied \cite{jin1997automatic,sadjadi20172016}. A typical pipeline is based on three main steps: \emph{i.a)} acoustic feature extraction from audio samples, \emph{i.b)} voice feature aggregation from the lower-level acoustic features by means of a speaker modeling stage, and \emph{ii)} a clustering technique on top of this feature-based representation. 

The voice features after phase \emph{i)} have been traditionally created based on Mel Frequency Cepstral Coefficient (MFCC) acoustic features modeled by a Gaussian Mixture Model (GMM) \cite{campbell2006support}, or i-vectors \cite{dehak2009support,lee2014clustering}. More recently, with the rise of deep learning, the community is moving towards learned features instead of hand-crafted ones, as surveyed by Richardson et al. \cite{richardson2015deep}. Recent examples of deep-feature representations for SI, SV, and SC problems come for example from Lukic et al. \cite{MLSP2017}, after Convolutional neural networks (CNN) have been introduced in the speech processing field by LeCun et al. already in the nineties \cite{lecun1995convolutional}. McLaren et al. used a CNN for speaker recognition in order to improve robustness to noisy speech \cite{mclaren2014application}. Chen et al. used a novel deep neural architecture to learn speaker specific characteristics directly from MFCC features \cite{chen2011learning}. Yella et al. exploited the capabilities of an artificial neural network of 3 layers to extract features directly from a hidden layer, which are used for speaker clustering \cite{yella2014artificial}. 


However advanced phase \emph{i)} has \FH{become} during the last years, the clustering phase \emph{ii)} still relies on traditional methodologies. For example, Khoury et al. demonstrated good results for speaker clustering using a hierarchical clustering algorithm \cite{khoury2014hierarchical}, while Kenny et al. report hierarchical clustering to be unsuitable for the speaker clustering stage in a speaker diarization system \cite{kenny2010diarization}. In \cite{shum2011exploiting} they performed clustering with K-means on dimensionality-reduced i-vectors which showed to work better than spectral clustering as noted in \cite{shum2012use}.

\begin{figure*}[!t]
\centering
\includegraphics[width=0.97\textwidth]{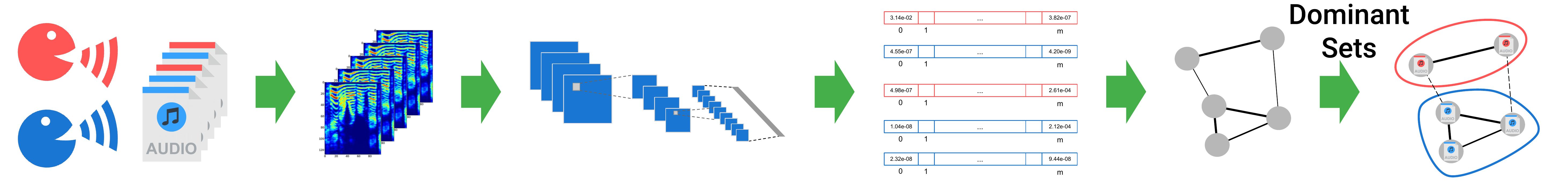}
\caption{Pipeline of the overall sequence of elaborations: voices $\rightarrow$ spectrograms $\rightarrow$ CNN $\rightarrow$ feature vectors $\rightarrow$ graph $\rightarrow$ Dominant Set $\rightarrow$ clusters.}
\label{img:pipeline}
\vspace{-0.4cm}
\end{figure*}




In this paper, we therefore improve the results of the speaker clustering task by first using state-of-art learned features and then, a different and more robust clustering algorithm, dominant sets (DS) \cite{pavan2007dominant}. The motivation driving the choice of dominant sets is the following: \emph{a)} no need for an a-priori number of clusters; \emph{b)} having a notion of compactness to be able to automatically detect clusters composed of noise; \emph{c)} for each cluster the centrality of each element is quantified (centroids emerge naturally in this context); and \emph{d)} extensive experimentations and the underlying theory prove a high robustness to noise \cite{pavan2007dominant}. All the aforementioned properties perfectly fit the SC problem.

The contribution of this paper is three-fold: first, we apply the dominant set method for the first time in the SC domain, outperforming the previous state of the art; second, it is the first time that the full TIMIT dataset \cite{timit:1986} is used for SC problems, making this paper a reference baseline in this context and on this dataset; third, we use for the first time \FH{a pre-trained VGGVox\footnote{\url{https://github.com/a-nagrani/VGGVox}} network to extract features for the TIMIT dataset}, obtaining good results and demonstrating the capability of this embedding.

The remainder of this paper is organized as follow: in Sec \ref{sec:method} the proposed method is explained in detail (with Sec \ref{sec:features} having explanations for the different feature extraction methods, and Sec \ref{sec:ds} having an introduction to the theoretical foundations of DS). In Sec \ref{sec:exp} the experiments that have been carried out are explained and in Sec \ref{sec:results} we discuss the results before drawing conclusions in Sec \ref{sec:conclusions} together with future perspectives.

\section{Speaker clustering with dominant sets}
\label{sec:method}

Our proposed approach, called \name\, is based on the two-phase schema (see Fig.\ref{img:pipeline}): the first part in which features are extracted from each utterance and the second one in which from this feature-based representation the dominant sets are extracted. In this section, the specific parts are explained.

\subsection{Features extraction}
\label{sec:features}
We use two different feature extraction methods in this work that we call \thilofeat\ (derived from embeddings based on the TIMIT dataset), and \zissfeat\ (based on a model trained on VoxCeleb \cite{Nagrani17}):

\subsubsection{\thilofeat\ features}
Features are extracted from the CNN\footnote{\url{https://github.com/stdm/ZHAW_deep_voice}} described in detail by Lukic et al. \cite{MLSP2016}, specifically from the dense layer L7 therein. The network has been trained on 590 speakers of the TIMIT database that have been fed to the net as spectrograms derived from the corresponding utterances, and yields 1,000-dimensional feature vectors.

\subsubsection{\zissfeat\ features}
Features are extracted from the published VGGVox model trained on the 100,000 utterances of the VoxCeleb dataset\cite{Nagrani17}. Since the domain of VoxCeleb and TIMIT are similar, we expect to have good performances on the latter. VGGVox is based on the VGG-M convolutional architecture \cite{chatfield2014return} which was previously used for image data, adapted for spectrogram input. We get 1,024-dimensional features from the FC7 layer as in the original publication. 


\subsection{Dominant Set clustering}
\label{sec:ds}
Dominant set clustering is a graph-based method that generalizes the problem of finding a maximal clique to edge-weighted graphs. A natural application of this method is for partitioning (clustering) a graph into disjoint sets. In this framework, a dataset is modeled as an undirected edge-weighted graph $G=(V,E,w)$ with no self loops, in which the nodes $V$ are the items of the dataset (represented by feature vectors). The edges $E \subseteq V \times V$ are the pairwise relations between nodes and their weight function $\omega: E \rightarrow \mathbb{R}_{\geq 0}$ calculates pairwise similarities. The $n \times n$ symmetric adjacency matrix $A=(a_{ij})$ is employed to summarize $G$:
\begin{equation}
a_{ij}= \left\{
  \begin{array}{l l}
    w(i,j) & \quad \textrm{if $(i,j) \in E$}\\
    0 & \quad \textrm{otherwise.}
  \end{array} \right. \nonumber \\
\end{equation}

Typically with every clustering method two properties shall hold: the \textit{intra-cluster} homogeneity is high  while \textit{inter-cluster homogeneity} is low. These two properties are important in order to separate and group objects in the best possible way. They are directly reflected in the combinatorial formulation of DS (see \cite{pavan2007dominant} for the details). 
Pavan and Pelillo propose an intriguing connection between clusters, dominant sets and local solutions of the following quadratic problem \cite{pavan2007dominant}:
\begin{eqnarray}
    \text{maximize }   & \textbf{x}^TA\textbf{x} \label{eqn:xTAx} \\
    \text{subject to } & \textbf{x} \in \bigtriangleup^n \nonumber
\end{eqnarray}
where $A$ is the similarity matrix of the graph and $\textbf{x}$ is the so-called \emph{characteristic vector} which lies in the n-dimensional simplex $\bigtriangleup^n$, that is, $\left(\sum_i{\textbf{x}_i = 1} , \forall i \mbox{  } x_i \geq 0\right)$. \FH{The components of vector $\textbf{x}$ represent the likelihood of each element to belong to the cluster, the higher the score the higher the chance of being part of it.}
If \textbf{x} is a strict local solution of (\ref{eqn:xTAx}) then its support $\sigma(\textbf{x})=\{ i \in V | x_i >0 \}$ is a dominant set.

In order to extract a DS, a local solution of (\ref{eqn:xTAx}) must be found. A method to solve this problem is to use a result from evolutionary game theory \cite{weibull1997evolutionary} known as \textit{replicator dynamic} (RD) (see Eq. \ref{eqn:repdyn}). 

\begin{equation}
x_i(t+1)=x_i(t)\frac{(A\textbf{x}(t))_i}{\textbf{x}(t)^TA\textbf{x}(t)} \label{eqn:repdyn}
\end{equation}
RD is a dynamical system that operates a selection process over the components of the vector \textbf{x}. At convergence of Eq. \ref{eqn:xTAx}  $(||\left(\textbf{x}(t)-\textbf{x}(t+1)\right)||_2 \leq \epsilon)$, certain components will emerge ($x_i > 0$) while others will get extinct ($x_i=0$). 
In practical cases, if these last components of $x$ are not exactly equal to zero then a thresholding ($x_i > \theta$) is performed. The convergence of the process is guaranteed if the matrix $A$ is non-negative and symmetric. The dynamical system starts at the barycenter of the simplex and its components are updated using Eq. \ref{eqn:repdyn}.

Deciding upon a cutoff threshold $\theta$ is not obvious. Instead of using a predefined value, we prefer to employ the approach proposed by Vascon et al. \cite{Vascon2013, dodero2013automatic}. The parameter is computed based on the following idea: it decides the minimum degree of participation of an element to a cluster and is relative to the participation of the centroid. The support for each dominant set becomes $\sigma(\textbf{x})=\{ i \in V | x_i > \theta * max(x) \}$ with $\theta \in \left[0,1\right)$ (see Sec. \ref{sec:res-sensitivity} for sensitivity analysis on the parameters).

At each iteration a dominant set is extracted and its subsets of nodes are removed from the graph (this is \FH{called} \emph{peeling-off} strategy). The process iterates on the remaining nodes until all are assigned to a cluster.

\subsection{Similarity measure}
\label{sec:simfunc}
To compute weights on edges of graph $G$ we use the \emph{cosine} distance to construct a similarity function. The cosine distance has been chosen because it showed good performance on SC tasks \cite{khoury2014hierarchical,shum2011exploiting,Nagrani17}. Given two utterances and their $m$-dimensional feature vectors $f_i$ and $f_j$, we apply the following function:
\begin{equation}\label{eqn:sim_funct}
\omega(f_i,f_j) = exp\left\lbrace-\frac{d(f_i,f_j)}{\sigma}\right\rbrace
\end{equation}
where $d$ is the cosine distance between given \FH{features}, \FH{and $\sigma$ is the similarity scaling parameter}.

Setting the parameter $\sigma$ is often problematic and typically requires a grid search over a range of plausible values or a cross-validation. In this work, we decided to use a principle heuristic from spectral clustering \cite{perona} which proved to work well also in other works \cite{tripodi2016context,ZemeneAP17}. Based on \cite{perona} and \cite{ZemeneAP17} we tested a local scaling parameter $\sigma_i$ for each utterance to be clustered. This means that in (\ref{eqn:sim_funct}) our parameter $\sigma = \sigma_i\sigma_j$ depends on local neighborhoods of given features $f_i$ and $f_j$ and it is determined as follows:
\begin{equation}
\sigma_i = \frac{1}{|N_i|}\sum_{k \in N_i}d(f_i, f_k)
\end{equation}
where $N_i$ represents the nearest neighborhood of element $i$. In our experiments we used $|N_i|=7$ as in \cite{ZemeneAP17}.

\subsection{Cluster labeling}
\label{sec:cluster_labeling}
Once all dominant sets are extracted, the final step is to label each partition such that each speaker is in one-to-one correspondence with a cluster. The labels of the data are then used to perform the assignment. We tested two approaches for cluster labeling:

\subsubsection{Max} a prototype selection method which assigns cluster labels using the class of the element with \emph{maximum} participation in the characteristic vector \cite{Vascon2013}. Labels are unique, and in case 2 different clusters share their labels, the latter one is considered completely mistaken, increasing error in the evaluation.

\subsubsection{Hungarian} with this approach, each cluster is labeled using the Munkres (aka Hungarian) method \cite{hungarian}. The cost $c_{i,j}$ of assigning a cluster $i$ to a particular label $j$ is computed as the number of elements of class $j$ in the cluster $i$. Since the method minimizes the total cost of assignments, the value of $c_{i,j}$ is changed to $\hat{c}_{i,j}=\max(c)-c_{i,j}$. This turns the minimization problem to a maximization one, where $\max(c)$ is the maximum cost over all the assignments.

\section{Experiments}
\label{sec:exp}

\subsection{Datasets \& data preparation}
We evaluate our method on the TIMIT dataset, presented as \emph{TIMIT Small} and \emph{TIMIT Full} 
(see Table \ref{tbl:datasets}). The dataset is composed of 6,300 phrases (10 phrases per speaker), spoken by 438 males (70\%) and 192 females (30\%). Speakers coming from 8 different regions and having different dialects. The phrases of each speaker have been divided into 2 parts in accordance with previous research \cite{Thilo2009,MLSP2016,MLSP2017}. In our experimentation we used the same 40 speakers dataset as reported by these earlier attempts (here called TIMIT Small), and the full TIMIT set composed by 630 speakers. \FH{Note that \emph{TIMIT Small} is disjoint with the training set of CNN-T.} 
\FH{This dataset is suited to our work because we are not dealing with noise, segmentation or similar diarization problems.}


\begin{table}[]
\centering
\label{tbl:datasets}
\caption{Datasets used in this paper.}
\begin{tabular}{l|c|c|c|c}
              & Acronym & \#POIs & \#Utt/POI & Utterances \\ \hline
TIMIT Small \cite{Thilo2009} & TimitS & 40     & 2         & 80        \\
TIMIT Full \cite{timit:1986} & TimitF & 630    & 2         & 1260      \\
\end{tabular}
\vspace{-0.4cm}
\end{table}
\subsection{Comparison to other methods}
The proposed method has been compared with the state of the art \cite{MLSP2017,MLSP2016,Thilo2009} and with other clustering techniques like  spectral clustering (SP), k-means (KM) and hierarchical clustering (HC). Given the fact that our proposed method is completely unsupervised (in particular, there is no knowledge a-priori of the number of clusters), we tested some heuristics to estimate $k$ also for the aforementioned algorithms. Specifically, the Eigengap heuristic \cite{von2007tutorial} and the number of clusters found by our method are used. 
Moreover, we chose affinity propagation (AP) \cite{frey2007clustering} and HDBSCAN \cite{hdbscan} because they do not require an a-priori $k$. 
In order to fairly compare our method, we tested them with the best settings. Specifically for HC and KM, cosine distance was the best choice, while for SP we used RBF kernel with $\gamma$ parameter found through an extensive grid search. The cut on HC has been set such that the error is minimized as in \cite{MLSP2016}. 
\FH{For AP we used the same similarity measure of \name\ while for HDBSCAN the Euclidean distance and minimum number of points per cluster equal to 2 were used.}


\subsection{Evaluation criteria} 
\label{sec:evaluationmetrics}
To evaluate the clustering quality we used three distinct metrics: the \emph{misclassification rate} (MR) \cite{Kotti}, the \emph{adjusted RAND index} (ARI) \cite{hubert1985comparing} and the \emph{average cluster purity} (ACP) \cite{solomonoff}. The usage of different metrics is important because each of them gives a different perspective on results: MR quantifies how many labels of speakers are inferred correctly from clusters while ARI and ACP are measures of grouping/separation performance on utterances. 

Formally, given a one-to-one mapping between clusters and labels (see Sec \ref{sec:cluster_labeling}), MR is defined as $MR = \frac{1}{N}\sum^{N_s}_{j=1}e_j$
where $N$ is the total number of audio segments to cluster, $N_s$ the number of speakers, and $e_j$ the number of segments of speaker $j$ classified incorrectly. 
\textit{Cluster purity} is a measure to determine how pure clusters are. If a cluster is composed of utterances belonging to the same speaker, then it is completely pure, otherwise (i.e., other speakers are in that cluster, too) purity decreases. 
Formally, average cluster purity is defined as:
\vspace{-0.4cm}\begin{displaymath}
acp = \frac{1}{N} \sum_{i=1}^{N_c} p_{i.}\cdot n_{i.}\mbox{ , where } p_{i.} = \sum_{j=1}^{N_s}n_{ij}^2 / n_{i.}^2
\end{displaymath}
\FH{$N_c$ is the number of clusters, $n_{ij}$ utterances in cluster $i$ spoken by speaker $j$ and $n_{i.}$ is the size of cluster $i$.}
The ARI finally is the normalized version of RAND index \cite{rand1971objective}, with maximum value~1 for perfectly assigned clusters with respect to the expected ones. 

\subsection{Experimental setup}

Our proposed method is evaluated in experiments composed as follows: given a set of audio utterances, features are extracted following one of the methods in Sec \ref{sec:features} and the affinity matrix is computed as in Sec \ref{sec:simfunc}. Subsequently, the DS are found on top of this graph-based representation. Labeling is performed on each cluster following the methodology proposed in Sec \ref{sec:cluster_labeling}. The goodness of clusters are then evaluated using the metrics in Sec \ref{sec:evaluationmetrics}. The summarized results are reported in Tables \ref{tbl:timitsmall_res} and \ref{tbl:timitfull_res} and discussed in the next section. The hyper parameters for all experiments are set to $\theta=0.1$ and $\varepsilon=1e-6$.




\section{Results}\label{sec:results}
In this section, the results of a series of analyses are reported, followed by an overall discussion.
\subsection{Initialization of \texttt{k} in the competitors}
DS does not need an a-priori number of clusters, while the supervised competitors do. In order to make a fair comparison with standard approaches (HC, KM and SP), we used as $k$: the correct number of clusters to be found (symbol $\diamondsuit$ in tables \ref{tbl:timitsmall_res} and \ref{tbl:timitfull_res}), the number of clusters found by DS (symbol $k$ in the tables) and the number of clusters estimated with \emph{Eigengap} (symbol \# in tables). 

Experimental results show that even when the correct number of clusters is provided, \name\ still achieves more desirable results (see tables). This means that not only our method is able to recover a number of clusters close to the right one, but also that it is able to extract much more correct partitions. And when the number of clusters found by DS is given to the other methods, results obtained are plausible, showing that our method is able to grasp a good number of clusters while with standard heuristics the performance drops strongly.



\subsection{Analysis of different feature extraction methods}
In the next experiments, we tested the two CNN-based features, \thilofeat\ and \zissfeat. Both provide good features in term of capacity to discriminate speakers. With the \thilofeat\ features, the performance of our method saturates on TIMIT Small (see last rows of Table \ref{tbl:timitsmall_res}) and reaches almost perfect performances on TIMIT Full (see last rows of Table \ref{tbl:timitfull_res}). This is mainly explained by the fact that the network used to extract the \thilofeat\ embeddings has been trained in using the remaining 590 of the 630 TMIT speakers \cite{MLSP2016}, thus \FH{biasedly} performing well on the entire dataset.

Surprisingly, features obtained from VGGVox are so generic that they allow almost the same performances for \name. This \FH{approach} is also beneficial for competitors, and in fact all of them have better performances in term of MR/ARI/ACP with \zissfeat\ features rather than \thilofeat\ ones (except for KM).

\subsection{Cluster labeling}
We tested two methods for labeling clusters for our approach (see \emph{Max} and \emph{Hungarian} in Sec \ref{sec:cluster_labeling}), while for all the other competitors we used only the \emph{Hungarian} algorithm since \emph{Max} is a peculiarity of DS. Under all conditions and datasets both labeling methods perform the same (see last rows of results where \emph{Max} = \name $\star$, \emph{Hungarian} = \name). Labeling with \emph{Max} method comes for free directly from DS theory, while applying the Hungarian method has its computational cost.

\subsection{Metrics comparison}
The three metrics (MR, ARI and ACP) are important to be analyzed in conjunction because they capture different aspects of the result. Having the lowest MR in the final results in both datasets emphasize the fact that we are correctly labeling clusters and that the number of miss-classified samples is extremely low. On the other side, reaching the highest value in ARI shows that our method obtains a good partitioning of the data with respect to the expected clusters. Furthermore, having the higher ACP confirms that clusters extracted with \name\ are mainly composed by utterances from the same speaker. 

The proposed method reaches best scores on all metrics simultaneously. Indeed, other methods reach similar performances, in particular on TIMIT Small (like HC, AP), but none of them work as well as our in the most complex experimental setting used, TIMIT Full with VGGVox features (where no knowledge of the actual voices to be clustered could possibly enter the features and thus the clustering system). 


\begin{table}[t!]
\footnotesize
\centering
\caption{Clustering results on the Timit Small dataset.}
\label{tbl:timitsmall_res}
\hspace*{-0.4cm}\begin{tabular}{l|ccc|ccc}
\multirow{2}{*}{\begin{tabular}[c]{@{}l@{}}SMALL \\ TIMIT\end{tabular}} & \multicolumn{3}{c|}{\thilofeat\ Features} & \multicolumn{3}{c}{\zissfeat\ Features} \\
                                                                        & MR $\downarrow$        & ARI $\uparrow$        & ACP $\uparrow$        & MR $\downarrow$       & ARI $\uparrow$      & ACP $\uparrow$      \\ \hline
HC $\diamondsuit$                                                                     
& 0.0250 & 0.9259 & 0.9667 & 0.0000 & 1.0000 & 1.0000 \\
HC \cite{MLSP2016}                                                                     & 0.0500 & - & -  & - & - & - \\
{\scriptsize Adadelta 20k\cite{MLSP2017}}                                                                     & 0.0500 & - & -  & - & - & - \\
{\scriptsize Adadelta 30k\cite{MLSP2017}}                                                                     & 0.0500 & - & -  & - & - & - \\
$\nu$-SVM \cite{Thilo2009} 
& 0.0600 & - & -  & - & - & - \\
{\scriptsize GMM/MFCC\cite{Thilo2009}} 
& 0.1300 & - & -  & - & - & - \\
SP $\diamondsuit$                                                                       
& 0.0750 & 0.8422 & 0.9500 & 0.0000 & 1.0000 & 1.0000 \\
KM $\diamondsuit$                                                                    
& 0.0250 & 0.9259 & 0.9667 & 0.0375 & 0.9390 & 0.9750 \\
\hline
HC k                                                                       
& 0.0250 & 0.9259 & 0.9667 & 0.0000 & 1.0000 & 1.0000 \\ 
SP k                                                                        
& 0.0750 & 0.8422 & 0.9500 & 0.0000 & 1.0000 & 1.0000 \\ 
KM k                                                                       
& 0.0250 & 0.9259 & 0.9667 & 0.0375 & 0.9390 & 0.9750 \\ 
\hline
HC \#                                                                       
& 0.4500 & 0.4234 & 0.5500 & 0.6750 & 0.2466 & 0.3250 \\
SP \#                                                                       
& 0.4500 & 0.0827 & 0.5500 & 0.6750 & 0.1751 & 0.3038 \\ 
KM \#                                                                       
& 0.4500 & 0.3543 & 0.5267 & 0.6750 & 0.1746 & 0.3193 \\ 
\hline
\hline
AP
& 0.0500 & 0.8951 & 0.9416 & 0.0000 & 1.0000 & 1.0000 \\
HDBS
& 0.1000 & 0.8056 & 0.8833 & 0.0750 & 0.8422 & 0.9083 \\
\hline
\name                                                                    
& \textbf{0.0000} & \textbf{1.0000} & \textbf{1.0000} & \textbf{0.0000} & \textbf{1.0000} & \textbf{1.0000} \\
\name $\star$                                                                    & \textbf{0.0000} & \textbf{1.0000} & \textbf{1.0000} & \textbf{0.0000} & \textbf{1.0000} & \textbf{1.0000}        
\end{tabular}
\end{table}


\subsection{Sensitivity analysis}
\label{sec:res-sensitivity}
\begin{figure}[!ht]
\centering
\includegraphics[width=0.49\textwidth]{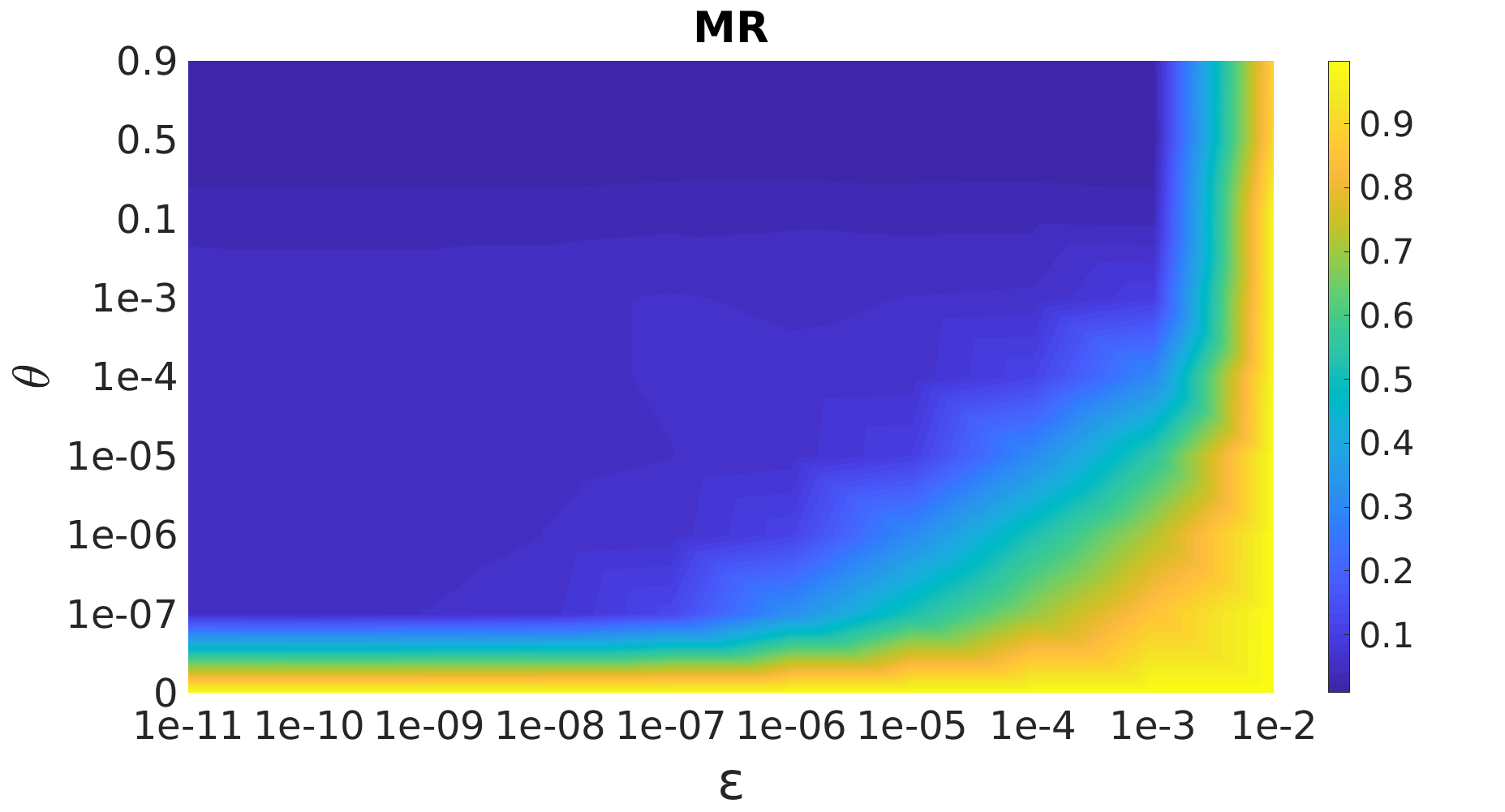}
\caption{Sensitivity of $\varepsilon$ and $\theta$ with respect to the MR measure.} 
\label{fig:mr_sensitivity}
\end{figure}

\begin{figure}[!ht]
\centering
\includegraphics[trim={0cm 0cm 1cm 0cm},clip,width=0.49\textwidth]{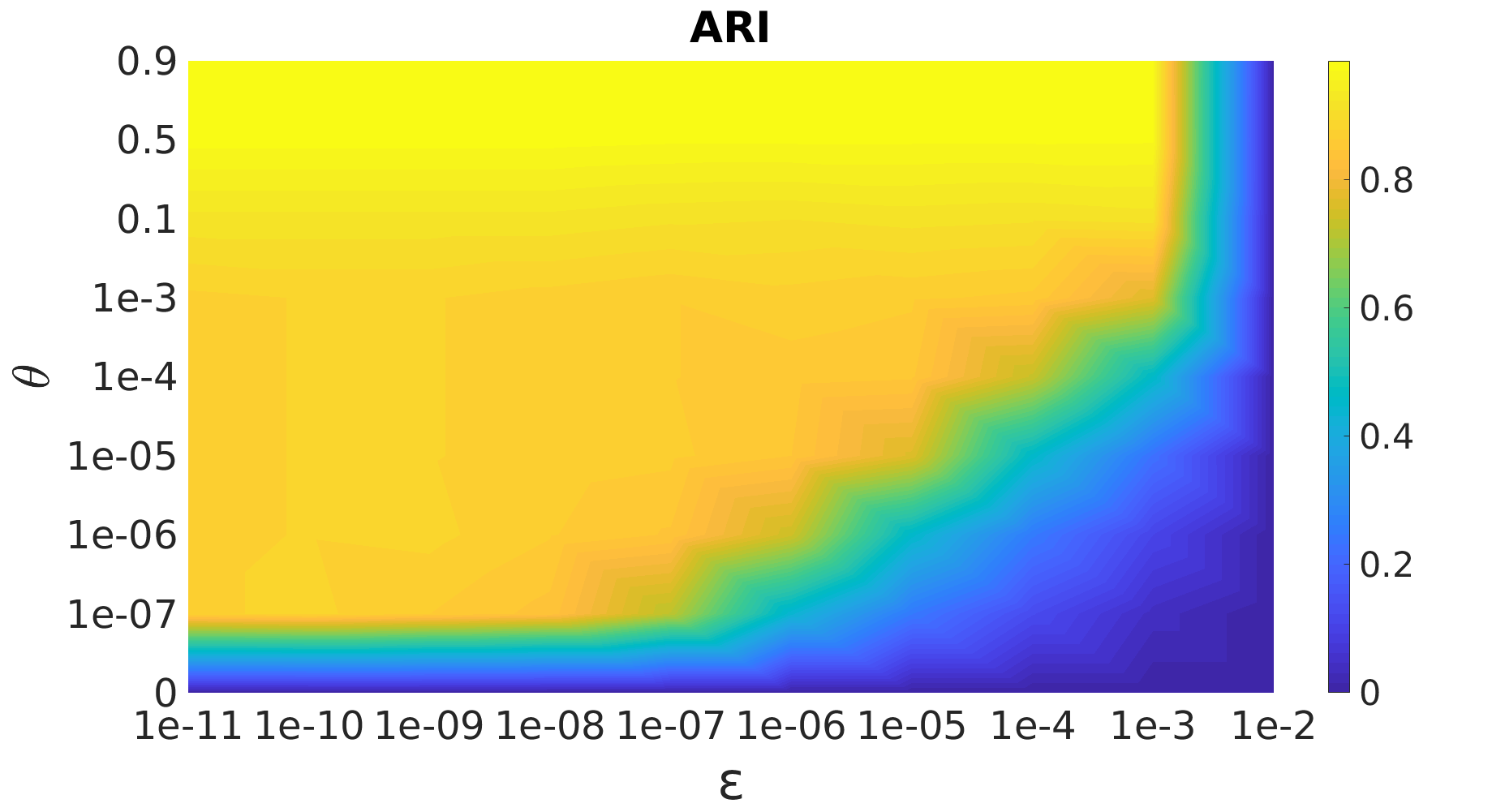}
\caption{Sensitivity of $\varepsilon$ and $\theta$ with respect to the ARI measure.}
\label{fig:ari_sensitivity}
\vspace{-0.4cm}
\end{figure}

Finally, we report the results of a sensitivity analysis on the two free-parameters of our method under two metrics (see Fig \ref{fig:mr_sensitivity} and \ref{fig:ari_sensitivity}),
the precision $\varepsilon$ of Replicator Dynamics (see Eq. \ref{eqn:repdyn}) and the relative cut-off $\theta$ (see Sec. \ref{sec:ds}). The analysis has been carried out on TIMIT Full with VGGVox features because under \FH{this setting} a certain amount of variability on results is observed, which made this analysis interesting. 
The search space for the parameters is as follows: $\theta \in \left[0.0,\ 0.9995\right]$ and $\varepsilon \in \left[1e-11,\  1e-2\right]$. The choice has been made on these extremal points for the following reasons: a low value, e.g. $\theta = 0.0005$, means that a point belongs to a cluster if and only if its level of participation in the cluster with respect to the centroid is at least $\theta\times$centrality of the centroid. Instead, $\theta = 0.9995$ means that the centroid and the sample must be almost exactly the same. In the first case we are creating clusters which span widely in terms of similarities of its elements, while in the latter case we create clusters composed by very similar elements. Regarding the parameter $\varepsilon$, when it is set to $1e-11$, it requires that two successive steps in Eq. \ref{eqn:repdyn} are very close to each other while in the case $1e-2$ we allow for a coarse equilibrium point.

Changes in both variables showed that the area in which the performances are stable is large (see the blue area in Fig \ref{fig:mr_sensitivity}
and yellow area in Fig \ref{fig:ari_sensitivity}). Only when extremal values of parameters are used the performances drops. The best parameter choice (\thilofeat: $\theta=0.15$, $\varepsilon=1e-6$; \zissfeat: $\theta=0.67$, $\varepsilon=1e-7$) is shown in Table \ref{tbl:timitfull_res} as \name best.

\begin{table}[t]
\footnotesize
\centering
\caption{Clustering results on the Timit Full dataset.}
\label{tbl:timitfull_res}
\begin{tabular}{l|ccc|ccc}
\multirow{2}{*}{\begin{tabular}[c]{@{}l@{}}FULL \\ TIMIT\end{tabular}} & \multicolumn{3}{c|}{\thilofeat\ Features} & \multicolumn{3}{c}{\zissfeat\ Features} \\
                                                                        & MR $\downarrow$        & ARI $\uparrow$        & ACP $\uparrow$        & MR $\downarrow$       & ARI $\uparrow$      & ACP $\uparrow$      \\ \hline
HC $\diamondsuit$                                                                       
& 0.0770 & 0.8341 & 0.9283 & 0.0571 & 0.8809 & 0.9484 \\
SP $\diamondsuit$                                                                    
& 0.2294 & 0.0432 & 0.8355 & 0.0675 & 0.5721 & 0.9488 \\ 
KM $\diamondsuit$                                                                       
& 0.1071 & 0.7752 & 0.9071 & 0.1286 & 0.6982 & 0.8730  \\

\hline
HC k                                                                       
& 0.0762 & 0.8343 & 0.9280 & 0.0706 & 0.8502 & 0.9295 \\ 
SP k                                                                       
& 0.2341 & 0.0421 & 0.8332 & 0.0635 & 0.4386 & 0.9427 \\ 
KM k                                                                       
& 0.1079 & 0.7682 & 0.9007 & 0.1429 & 0.6646 & 0.8485 \\ 
\hline
HC \#                                                                       
& 0.9921 & 0.0050 & 0.0079 & 0.9984 & 0.0000 & 0.0016 \\ 
SP \#                                                                       
& 0.9921 & 0.0003 & 0.0075 & 0.9984 & 0.0000 & 0.0016 \\ 
KM \#                                                                       
& 0.9921 & 0.0052 & 0.0076 & 0.9984 & 0.0000 & 0.0016 \\ 
\hline
\hline
AP
& 0.0753 & 0.8330 & 0.9030 & 0.1396 & 0.7127 & 0.8222 \\
HDBS
& 0.1825 & 0.6214 & 0.7825 & 0.3000 & 0.4112 & 0.6527 \\
\hline
\name                                                                      
& \textbf{0.0048} & \textbf{0.9897} & \textbf{0.9947} & \textbf{0.0349} & \textbf{0.9167} & \textbf{0.9578}       \\
\name $\star$ & \textbf{0.0048} & \textbf{0.9897} & \textbf{0.9947} & \textbf{0.0349} & \textbf{0.9167} & \textbf{0.9578} \\
\name  best & \textbf{0.0032} & \textbf{0.9929} & \textbf{0.9966} & \textbf{0.0024} & \textbf{0.9944} & \textbf{0.9974} 

\end{tabular}
\end{table}



\vspace{-0.3cm}

\subsection{Overall discussion}
From a global perspective we can say that the proposed \name\ method performs better than the alternatives on the used datasets, outperforming the state-of-the-art and showing a more adaptive response also \FH{with a pre-trained model on a different dataset}. In particular, this \FH{is} evident in TIMIT Full, where better performances than competitors are achieved even when they are given the right number of clusters to be found. It is worth to note that our clustering method has only two parameters to set, which are both very insensitive to variation as shown in the sensitivity analysis.


Interesting to note, an analysis of misclassified speakers shows that if a speaker is wrongly clustered by DS, it is also wrongly clustered by all other methods. This gives rise to the assumption that in these cases the extracted features may be the reason for the error rather than the clustering approach used. 



\section{Conclusions}
\label{sec:conclusions}
In this paper, we have proposed a novel pipeline for speaker clustering. The proposed method is based on the dominant set clustering algorithm which has been applied to this domain for the first time. It outperforms the previous state of the art and other clustering techniques. 

We proposed a method which is almost parameter-less --  the two free parameters do not affect too much the results, testifying to its stability. Moreover, we successfully  \FH{used a pre-trained CNN model on a} different dataset and report reasonable speaker clustering performance on the TIMIT Full dataset for the first time (after the $99.84$\% MR reported by Stadelmann and Freisleben using a classical pipeline \cite{Thilo2009}). \FH{Now that we reached a good starting point with noise-free utterances we can start considering more complex datasets with their relatively more challenging tasks (noise, segmentation, cross-talk etc.)}. Future work also includes improving the features using the siamese network proposed by Nagrani et al. \cite{Nagrani17} to extract similarities directly.\\

Code available at \url{https://github.com/feliksh/SCDS}

\section*{Acknowledgment}
The authors would like to thank Y. Lukic for providing the MLSP features and J. Salamone for contributing to first results.
\bibliographystyle{IEEEtran}
\bibliography{biblio}

\begin{thebibliography}{10}
\providecommand{\url}[1]{#1}
\csname url@samestyle\endcsname
\providecommand{\newblock}{\relax}
\providecommand{\bibinfo}[2]{#2}
\providecommand{\BIBentrySTDinterwordspacing}{\spaceskip=0pt\relax}
\providecommand{\BIBentryALTinterwordstretchfactor}{4}
\providecommand{\BIBentryALTinterwordspacing}{\spaceskip=\fontdimen2\font plus
\BIBentryALTinterwordstretchfactor\fontdimen3\font minus
  \fontdimen4\font\relax}
\providecommand{\BIBforeignlanguage}[2]{{%
\expandafter\ifx\csname l@#1\endcsname\relax
\typeout{** WARNING: IEEEtran.bst: No hyphenation pattern has been}%
\typeout{** loaded for the language `#1'. Using the pattern for}%
\typeout{** the default language instead.}%
\else
\language=\csname l@#1\endcsname
\fi
#2}}
\providecommand{\BIBdecl}{\relax}
\BIBdecl

\bibitem{beigi2011fundamentals}
H.~Beigi, \emph{Fundamentals of speaker recognition}.\hskip 1em plus 0.5em
  minus 0.4em\relax Springer Science \& Business Media, 2011.

\bibitem{anguera2012speaker}
X.~Anguera, S.~Bozonnet, N.~Evans, C.~Fredouille, G.~Friedland, and O.~Vinyals,
  ``Speaker diarization: A review of recent research,'' \emph{IEEE Transactions
  on Audio, Speech, and Language Processing}, vol.~20, no.~2, pp. 356--370,
  2012.

\bibitem{ajmera2003robust}
J.~Ajmera and C.~Wooters, ``A robust speaker clustering algorithm,'' in
  \emph{IEEE Workshop on Automatic Speech Recognition and Understanding}, 2003.

\bibitem{shum2012use}
S.~Shum, N.~Dehak, and J.~Glass, ``On the use of spectral and iterative methods
  for speaker diarization,'' in \emph{Annual Conference of the International
  Speech Communication Association}.\hskip 1em plus 0.5em minus 0.4em\relax
  ISCA, 2012, pp. 482--485.

\bibitem{jin1997automatic}
H.~Jin, F.~Kubala, and R.~Schwartz, ``Automatic speaker clustering,'' in
  \emph{Proceedings of the DARPA speech recognition workshop}, 1997, pp.
  108--111.

\bibitem{sadjadi20172016}
S.~O. Sadjadi, T.~Kheyrkhah, A.~Tong, C.~Greenberg, E.~S. Reynolds, L.~Mason,
  and J.~Hernandez-Cordero, ``The 2016 nist speaker recognition evaluation,''
  \emph{Interspeech}, pp. 1353--1357, 2017.

\bibitem{campbell2006support}
W.~M. Campbell, D.~E. Sturim, and D.~A. Reynolds, ``Support vector machines
  using gmm supervectors for speaker verification,'' \emph{IEEE Signal
  Processing Letters}, vol.~13, no.~5, pp. 308--311, 2006.

\bibitem{dehak2009support}
N.~Dehak, R.~Dehak, P.~Kenny, N.~Br{\"u}mmer, P.~Ouellet, and P.~Dumouchel,
  ``Support vector machines versus fast scoring in the low-dimensional total
  variability space for speaker verification,'' in \emph{Annual Conference of
  the International Speech Communication Association}, 2009.

\bibitem{lee2014clustering}
H.-S. Lee, Y.~Tsao, H.-M. Wang, and S.-K. Jeng, ``Clustering-based i-vector
  formulation for speaker recognition,'' in \emph{Annual Conference of the
  International Speech Communication Association}, 2014.

\bibitem{richardson2015deep}
F.~Richardson, D.~Reynolds, and N.~Dehak, ``Deep neural network approaches to
  speaker and language recognition,'' \emph{IEEE Signal Processing Letters},
  vol.~22, no.~10, pp. 1671--1675, 2015.

\bibitem{MLSP2017}
Y.~Lukic, C.~Vogt, O.~Durr, and T.~Stadelmann, ``Learning embeddings for
  speaker clustering based on voice equality,'' in \emph{IEEE International
  Workshop on Machine Learning for Signal Processing (MLSP)}, 2017, pp. 1--6.

\bibitem{lecun1995convolutional}
Y.~LeCun, Y.~Bengio \emph{et~al.}, ``Convolutional networks for images, speech,
  and time series,'' \emph{The Handbook of Brain Theory and Neural Networks},
  vol. 3361, no.~10, p. 1995, 1995.

\bibitem{mclaren2014application}
M.~McLaren, Y.~Lei, N.~Scheffer, and L.~Ferrer, ``Application of convolutional
  neural networks to speaker recognition in noisy conditions,'' in \emph{Annual
  Conference of the International Speech Communication Association}, 2014.

\bibitem{chen2011learning}
K.~Chen and A.~Salman, ``Learning speaker-specific characteristics with a deep
  neural architecture,'' \emph{IEEE Transactions on Neural Networks}, vol.~22,
  no.~11, pp. 1744--1756, 2011.

\bibitem{yella2014artificial}
S.~H. Yella, A.~Stolcke, and M.~Slaney, ``Artificial neural network features
  for speaker diarization,'' in \emph{IEEE Spoken Language Technology Workshop
  (SLT)}.\hskip 1em plus 0.5em minus 0.4em\relax IEEE, 2014, pp. 402--406.

\bibitem{khoury2014hierarchical}
E.~Khoury, L.~El~Shafey, M.~Ferras, and S.~Marcel, ``Hierarchical speaker
  clustering methods for the nist i-vector challenge,'' in \emph{Odyssey: The
  Speaker and Language Recognition Workshop.}, 2014.

\bibitem{kenny2010diarization}
P.~Kenny, D.~Reynolds, and F.~Castaldo, ``Diarization of telephone
  conversations using factor analysis,'' \emph{IEEE Journal of Selected Topics
  in Signal Processing}, vol.~4, no.~6, pp. 1059--1070, 2010.

\bibitem{shum2011exploiting}
S.~Shum, N.~Dehak, E.~Chuangsuwanich, D.~Reynolds, and J.~Glass, ``Exploiting
  intra-conversation variability for speaker diarization,'' in \emph{Annual
  Conference of the International Speech Communication Association}, 2011, pp.
  945--948.

\bibitem{pavan2007dominant}
M.~Pavan and M.~Pelillo, ``Dominant sets and pairwise clustering,'' \emph{IEEE
  Transactions on Pattern Analysis and Machine Intelligence}, vol.~29, no.~1,
  pp. 167--172, 2007.

\bibitem{timit:1986}
W.~M. Fisher, G.~R. Doddington, and K.~M. Goudie-Marshall, ``{The DARPA Speech
  Recognition Research Database: Specifications and Status},'' in
  \emph{Proceedings of DARPA Workshop on Speech Recognition}, 1986, pp. 93--99.

\bibitem{Nagrani17}
A.~Nagrani, J.~S. Chung, and A.~Zisserman, ``Voxceleb: a large-scale speaker
  identification dataset,'' in \emph{Interspeech}, 2017.

\bibitem{MLSP2016}
Y.~Lukic, C.~Vogt, O.~Dürr, and T.~Stadelmann, ``{Speaker Identification and
  Clustering Using Convolutional Neural Networks},'' in \emph{IEEE
  International Workshop on Machine Learning for Signal Processing (MLSP)},
  Sept 2016, pp. 1--6.

\bibitem{chatfield2014return}
K.~Chatfield, K.~Simonyan, A.~Vedaldi, and A.~Zisserman, ``Return of the devil
  in the details: Delving deep into convolutional nets,'' in \emph{British
  Machine Vision Conference (BMVC)}, 2014.

\bibitem{weibull1997evolutionary}
J.~W. Weibull, \emph{Evolutionary game theory}.\hskip 1em plus 0.5em minus
  0.4em\relax MIT press, 1997.

\bibitem{Vascon2013}
S.~Vascon, M.~Cristani, M.~Pelillo, and V.~Murino, ``Using dominant sets for
  k-nn prototype selection,'' in \emph{International Conference on Image
  Analysis and Processing (ICIAP)}, A.~Petrosino, Ed.\hskip 1em plus 0.5em
  minus 0.4em\relax Springer Berlin Heidelberg, 2013, pp. 131--140.

\bibitem{dodero2013automatic}
L.~Dodero, S.~Vascon, L.~Giancardo, A.~Gozzi, D.~Sona, and V.~Murino,
  ``Automatic white matter fiber clustering using dominant sets,'' in
  \emph{International Workshop on Pattern Recognition in Neuroimaging
  (PRNI)}.\hskip 1em plus 0.5em minus 0.4em\relax IEEE, 2013, pp. 216--219.

\bibitem{perona}
L.~Zelnik-Manor and P.~Perona, ``Self-tuning spectral clustering,'' in
  \emph{Advances in neural information processing systems (NIPS)}, 2005, pp.
  1601--1608.

\bibitem{tripodi2016context}
R.~Tripodi, S.~Vascon, and M.~Pelillo, ``Context aware nonnegative matrix
  factorization clustering,'' in \emph{IEEE International Conference on Pattern
  Recognition (ICPR)}.\hskip 1em plus 0.5em minus 0.4em\relax IEEE, 2016, pp.
  1719--1724.

\bibitem{ZemeneAP17}
\BIBentryALTinterwordspacing
E.~Zemene, L.~T. Alemu, and M.~Pelillo, ``Dominant sets for "constrained" image
  segmentation,'' \emph{CoRR}, vol. abs/1707.05309, 2017. [Online]. Available:
  \url{http://arxiv.org/abs/1707.05309}
\BIBentrySTDinterwordspacing

\bibitem{hungarian}
H.~W. Kuhn, ``The hungarian method for the assignment problem,'' \emph{Naval
  Research Logistics (NRL)}, vol.~2, no. 1-2, pp. 83--97, 1955.

\bibitem{Thilo2009}
T.~Stadelmann and B.~Freisleben, ``Unfolding speaker clustering potential: a
  biomimetic approach,'' in \emph{International Conference on Multimedia},
  2009, pp. 185--194.

\bibitem{von2007tutorial}
U.~Von~Luxburg, ``A tutorial on spectral clustering,'' \emph{Statistics and
  computing}, vol.~17, no.~4, pp. 395--416, 2007.

\bibitem{frey2007clustering}
B.~J. Frey and D.~Dueck, ``Clustering by passing messages between data
  points,'' \emph{science}, vol. 315, no. 5814, pp. 972--976, 2007.

\bibitem{hdbscan}
R.~J. G.~B. Campello, D.~Moulavi, and J.~Sander, ``Density-based clustering
  based on hierarchical density estimates,'' in \emph{Advances in Knowledge
  Discovery and Data Mining}, J.~Pei, V.~S. Tseng, L.~Cao, H.~Motoda, and
  G.~Xu, Eds.\hskip 1em plus 0.5em minus 0.4em\relax Berlin, Heidelberg:
  Springer Berlin Heidelberg, 2013, pp. 160--172.

\bibitem{Kotti}
M.~Kotti, V.~Moschou, and C.~Kotropoulos, ``Review: Speaker segmentation and
  clustering,'' \emph{Signal Process.}, vol.~88, no.~5, pp. 1091--1124, May
  2008.

\bibitem{hubert1985comparing}
L.~Hubert and P.~Arabie, ``Comparing partitions,'' \emph{Journal of
  Classification}, vol.~2, no.~1, pp. 193--218, 1985.

\bibitem{solomonoff}
A.~Solomonoff, A.~Mielke, M.~Schmidt, and H.~Gish, ``Clustering speakers by
  their voices,'' in \emph{IEEE International Conference on Acoustics, Speech
  and Signal Processing (ICASSP)}, 1998.

\bibitem{rand1971objective}
W.~M. Rand, ``Objective criteria for the evaluation of clustering methods,''
  \emph{Journal of the American Statistical association}, vol.~66, no. 336, pp.
  846--850, 1971.

\end{thebibliography}


\end{document}